\documentclass[journal,12pt,onecolumn,draftclsnofoot,]{IEEEtran}

\IEEEoverridecommandlockouts
\usepackage{cite}
\usepackage{amsmath,amssymb,amsfonts}
\usepackage{graphicx}
\usepackage{textcomp}
\usepackage{multirow}
\usepackage{soul}

\usepackage[colorlinks]{hyperref}

\usepackage[printonlyused,withpage]{acronym}

\usepackage{xcolor}
\usepackage{algorithm}
\usepackage{algpseudocode}

\usepackage{amsthm}
\usepackage{tabularx}

\usepackage{eurosym}
\usepackage[caption=false]{subfig}

\usepackage{hyperref}

\def\BibTeX{{\rm B\kern-.05em{\sc i\kern-.025em b}\kern-.08em
    T\kern-.1667em\lower.7ex\hbox{E}\kern-.125emX}}
\begin{document}

\title{Evaluation of voltage magnitude based unbalance metric for low voltage distribution networks
}

\author{\IEEEauthorblockN{Md Umar Hashmi, Arpan Koirala, Rickard Lundholm, Hakan Ergun, Dirk Van Hertem\\}
\IEEEauthorblockA{\textit{Electa-ESAT, KU Leuven \& EnergyVille},
Genk, Belgium\\}
\IEEEauthorblockA{ (mdumar.hashmi, arpan.koirala, rickard.lundholm, hakan.ergun, dirk.vanhertem)@kuleuven.be}
}

\maketitle
{}
\begin{abstract}
Voltage unbalance in distribution networks (DN) is expected to grow with increasing penetration of single-phase distributed generation and single-phase loads such as electric vehicle chargers. 
Unbalance mitigation will be a significant concern as voltage unbalance leads to increased losses, reduced motor and inverter efficiency, and becomes a limiting factor for DN operation. 
The true definition of the unbalance metric needs phasor measurements of network voltage and current. 
However, such phasor measurements are generally not available in real life and as such approximate definitions are widely used due to their simplicity. 
This work aims to compare the true voltage unbalance definition and approximate unbalance metrics derived from phase voltage magnitude, as phase voltage magnitudes are commonly measured by digital metering infrastructure.
For the comparison, multi-period power flow simulations are performed for 161 Spanish distribution feeders with R/X ratios varying from 2.87 to 14.68.
We observe that phase magnitude-based unbalance metrics reasonably approximate the true unbalance for higher R/X ratios with a varying load power factor in a DN. Furthermore, the approximate unbalance metrics slightly improve for a low DN power factor due to the increase in DN unbalance. 
Therefore, the phase magnitude-based unbalance metric can be utilized for approximating DN unbalance.
\end{abstract}


\tableofcontents

\section*{List of Acronyms}

\begin{table*}[!htbp]
\centering
\begin{tabular}{c c}
DG & Distributed generation \\
DN    & distribution network  \\
DSO    & Distribution system operator  \\
EV & Electric vehicle\\
EU & European Union \\
LV & Low voltage\\
NEMA & National Electrical Manufacturers Association \\
PVUR & Phase voltage unbalance rate\\
SM & Smart meter\\
VUF & Voltage unbalance factor    \\ 
\end{tabular}
\end{table*}

\pagebreak



\section{Introduction}
Distribution network (DN) monitoring and control are governed and constrained by the measurement infrastructure.
Electricity suppliers or distribution system operators (DSO) install smart meters (SM) at the consumer end. These SMs, apart from electricity billing and advanced local energy management, can also be utilized to mitigate the network congestion at the end consumer using the measurement of power injection, power factor, and nodal voltage \cite{eusmartmeter}.
The EU estimates that more than 225 million smart meters will be installed in Europe by the year 2024 \cite{eubenchmark}.
In addition to smart meters, street or feeder head measurement boxes are also installed in some countries. For Germany, a minimum of 30 kW load aggregation is crucial to ensure the privacy of consumers.
Such limitations on measurement infrastructure constrain monitoring, preventive action planning, and control of low voltage (LV) DN.


Although most DN are three-phase networks, most DN loads {in Europe} are single-phase. As an example, the demonstration feeders used within the  \href{https://euniversal.eu/}{EUniversal project} approximately 80\% of the customer are connected single phase \cite{euniversal}. { Distribution network unbalance is caused by single phase connection of distributed generation installations, as well as with unbalanced loads, e.g. single phase charging of electric vehicles \cite{von2001assessment, chua2012energy, ul2015impact}. Voltage unbalance is not a major concern in most DN today. However, in the future, increasing power consumption (or injection) of single phase connected electric vehicles and PV generation may result in congestion where an active unbalance correction can be a cost-effective alternative to  grid reinforcement.}
Voltage unbalance increases the power losses and limits the loading capability of distribution transformers \cite{bina2011three,ochoa2005evaluation}. With ad hoc deployment of renewables, the chances of DN unbalance increase as the share of distributed renewable energy installations and unbalanced single phase load increases \cite{shahnia2010sensitivity,hashmi2020towards}. 
There are various definitions for calculating the voltage unbalance factor (VUF) of a DN.
The true definition of VUF requires phasor measurements of three-phase voltages.
Such measurements are often not available for monitoring and control of a LV DN. 
Most DN measurement infrastructure accurately measures three-phase voltage and active power magnitudes, see Table \ref{tab:ref}. Typically, detailed phase measurements come from smaller targeted campaigns, as the smart meter infrastructure required is typically more expensive and more complicated to manage. 
Prior works \cite{douglass2016voltage} propose and apply approximate magnitude based VUF definitions.
Authors in \cite{girigoudar2020impact} compare unbalance indicators used in literature. They observe that minimizing DN losses could also lead to a reduction in DN imbalances. 
Alternatively, \cite{bajo2015voltage} proposes the use of phase voltage and current magnitudes for measuring DN imbalances. These indices are evaluated in \cite{douglass2016voltage} and more recently in \cite{yao2020mitigating, yao2020overcoming}. 

\begin{table*}[!htbp]
\small
\caption{\small{Available smart meter data}}
\centering
\begin{tabular}{c|c|p{45mm}|c|c|p{29mm}}
Ref &
  Period &
  Measurement infrastructure &
  \# of devices &
  Country  &
  Parameters \\ \hline \hline
\cite{data_ireland} &
  2009-2010 &
  Open source (OS) SMs &
  6445 &
  Ireland &
  kWh per 30 min \\
\cite{data_low_carbon_london} &
  2011-2014 &
  OS SM data with tariff information &
  5567 &
  UK &
  kWh per 30 min \\
  
\cite{data_pecan} &
  2011-2021 &
  OS residential consumption &
  507 &
  USA &
  $P$ per 1 sec \\
  
\cite{data_smart} &
  One day &
  OS high and low-resolution residential consumption &
  400 &
  USA &
  $3$ phase $P$ per 1 sec \\
  
\cite{data_aus} &
  2010-2013 &
 OS bi-directional meter with data on controllable loads &
  300 &
  Australia &
  kWh per 30 min \\
  
  
  \cite{al_khafaf_impact_2022} &
    2018-2020 &
  Bi-directional meter &
  \textgreater{}5000 &
  Australia &
  kWh per 30 min \\
  
  \cite{wang_phase_2016} &
  Aug-Oct 2015 &
  $3-\phi$ SM with SCADA system &
  1500 &
  USA &
  $3$ phase and LL $V$/hour \\
  
\cite{luan_smart_2015} &
  One day &
  $3-\phi$ measurements with GIS data &
  3000/2000/700 &
  Canada &
  $3$ phase $V$ and kWh \\
  
 \cite{data_austria} &
 2009-2010&
    $3-\phi$ measurements &
    30 &
    Austria &
    $3-\phi$  $V$, $P$, and $Q$
\end{tabular}
\label{tab:ref}
\end{table*}

The goal of this paper is to compare an approximation of the VUF using phase voltage amplitude measurements with the true VUF definition, and elaborates when and where the approximation is valid.
There are many immediate applications of this work.
Widespread use of approximate VUF calculations can be developed with the existing DN measurement infrastructure. This will accelerate the deployment of renewable energy sources on distribution feeders without active imbalance compensation where hard constraints on VUF are getting violated.



\textit{Contributions:}
{Firstly, the paper demonstrates that phase voltage magnitude-based approximate VUF has a high Pearson correlation with the true VUF. We also show that magnitude-based VUF is an under-estimate for cases where the  DN is fairly balanced.}
{Secondly, the analysis in the paper shows that the power factor of DN does not impact the correlation of the two VUF metrics. Furthermore, with a deteriorating power factor, the correlation of VUF metrics increases due to an increase in DN imbalance.}
Our key conclusion is that for a DN with high R/X value, magnitude based VUF metric can  approximate fairly well the true VUF. 


The paper is organized as follows. 
Section~\ref{sec:section2} presents the VUF definitions evaluated in this work.
Section~\ref{sec:section3} presents three numerical case studies, and
section~\ref{sec:section4} concludes the paper.

\section{Voltage unbalance definition and metrics}
\label{sec:section2}
{In this section, we detail the voltage unbalance definitions, and the proposed metrics used to compare these definitions.}

\subsection{Notation}
A DN feeder is indexed as $f\in\{1,...,F\}$, where $F$ denotes the total number of feeders. 
A feeder $f$ consists of $N_f$ a number of total nodes.
Each node $i \in \{1,..., N_f\}$ has two variables for each phase, i.e., voltage magnitude ($V^{\phi}_{f,i,t}$) and phase angle ($\theta^{\phi}_{f,i,t}$) at each time instance $t$ which are governed by feeder characteristics, power injection and load magnitude. The three-phase of the DN are denoted as $\phi \in \{a,b,c\}$.
Multi-period simulations are conducted for $t\in\{1,...,T\}$, where $T$ is the total number of time steps. 
The DN cumulative resistance and reactance are given as R and X, respectively.
For a parameter $K$, $\bar{K}$ is the mean value, and $|K|$ denotes its absolute value.

\subsection{Introduction to power system unbalance}
\begin{figure}[!htbp]
	\centering
	\includegraphics[width=5.5in]{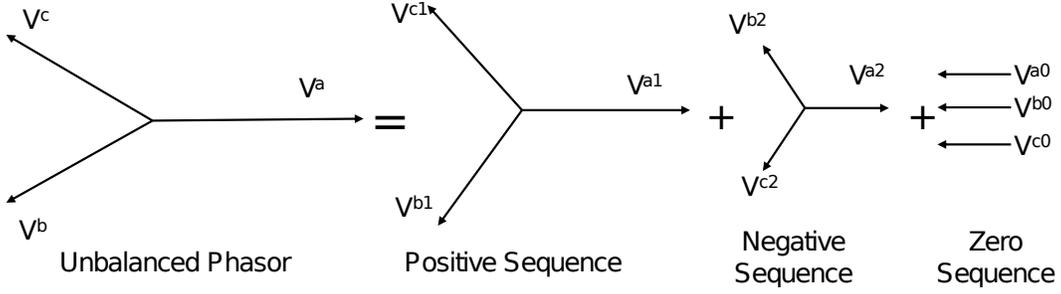}
	\caption{\small {Sequence component of an unbalanced voltage phasor \cite{hashmi2019optimization}} }
	\label{fig:seq}
\end{figure}

Three-phase voltage unbalance implies that the magnitude of the phase voltages and/or the phase angle between consecutive phases are not all equal \cite{r9beharrysingh2014phase}, \cite{r7bchydro}.
In 1918, Fortescue developed symmetrical components for representation of any set of unbalanced phasors \cite{fortescue1918method} (see Fig.~\ref{fig:seq}):
\begin{itemize}
	\item a direct or \textit{positive} sequence in order (abc),
	\item an inverse or \textit{negative} sequence in order (acb),
	\item a homopolar or \textit{zero} sequence system in the same direction.
\end{itemize}
The decomposition of unbalanced three-phase parameters into symmetrical components are shown in Fig.~\ref{fig:seq}.
Mathematically, the transformation of unbalanced three-phase voltages denoted as $\{V_{f,i,t}^a, V_{f,i,t}^b, V_{f,i,t}^c\}$ phasors are given as \cite{fortescue1918method, r2amberg2014tutorial}
\begin{equation}
{\begin{bmatrix}
V_{f,i,t}^{a0}\\V_{f,i,t}^{a1}\\V_{f,i,t}^{a2}\\
\end{bmatrix}} = \frac{1}{3} 
{\begin{bmatrix}
1 & 1&1\\
1& e^{\frac{j2\pi}{3}} & e^{\frac{-j2\pi}{3}}\\
1 & e^{\frac{-j2\pi}{3}} & e^{\frac{j2\pi}{3}}\\
\end{bmatrix}}
{\begin{bmatrix}
V_{f,i,t}^a\\ V_{f,i,t}^b\\V_{f,i,t}^c
\end{bmatrix}}
\end{equation}
where $e^{\frac{j2\pi}{3}} = 1 \angle 0^0$, $\{V_{f,i,t}^{a0}, V_{f,i,t}^{a1}, V_{f,i,t}^{a2}\}$ denote zero, positive and negative sequence components of three-phase voltages.

\subsection{Voltage unbalance definition}
Voltage unbalance can be measured using different metrics. Simplified definitions are proposed by 
National Electrical Manufacturers Association (NEMA) \cite{ieeestd}, and IEEE \cite{pillay2001definitions}.
In this work we compare two commonly used voltage unbalance metrics denoted as $U^{\text{true}}_{f,i,t}$ and $U^{\text{approx}}_{f,i,t}$.
Next, we provide the definitions of these unbalance metrics evaluated.

\subsubsection{True definition of VUF}
International Electrotechnical Commission or IEC defines voltage unbalance factor (VUF) as the ratio of the magnitudes of negative sequence over positive sequence, denoted as
\begin{equation}
U^{\text{true}}_{f,i,t} ~(\text{ in \%})= 100 \times \frac{|V^{a2}_{f,i,t}|}{|V^{a1}_{f,i,t}|}.
\label{eq:truedef}
\end{equation}
$U^{\text{true}}_{f,i,t}$ denotes the
VUF calculated using the true definition for feeder $f$, node $i$, and time $t$.
All other definitions of VUF are approximations of the true  definition of VUF \cite{pillay2001definitions}. 

\subsubsection{Approximate voltage imbalance indicator}
The approximate voltage imbalance indicator considers the phase voltage magnitude and ignores phase angle of voltage phasors. The main advantage is that it can be directly applied with most measurement infrastructure in DN, see Tab. \ref{tab:ref}.

Nodal voltage imbalance is defined using phase voltage unbalance rate (PVUR).
PVUR for feeder $f$, node $i$, at time $t$ and phase $\phi$ is denoted as 
\begin{equation}
    (\text{PVUR})_{f,i,t}^{\phi} =V_{f,i,t}^{\phi}/ \bar{V}_{f,i,t},
\end{equation}
where $\bar{V}_{f,i,t}=\frac{1}{3} \sum_{\phi \in \{a,b,c\}} V_{f,i,t}^{\phi}$ denotes the mean voltage at node $i$, time $t$ and for feeder $f$.
The PVUR for feeder $f$, node $i$ and time $t$ is given as
\begin{equation}
    \text{PVUR}_{f,i,t} = \max_{\phi \in \{A,B,C\}}\{ |\text{PVUR}_{f,i,t}^{\phi}| \}.
\end{equation}

Normalized voltage imbalance at node $i$ at time $t$ is given as
\begin{equation}
    U_{{f,i,t}}^{\text{approx}} = 1 -  \text{PVUR}_{f,i,t},
    \label{eq:approxdef}
\end{equation}
Next, we define the metrics for comparing the unbalance of DN.

\subsection{Metrics for comparing unbalance definitions}
The unbalance indicators need to capture the variability of DN voltage unbalance. The variation may not be equally matching in magnitude.
In order to make a meaningful comparison, we utilize correlation as a measure. 
For comparing the VUF definitions, we assume $U^{\text{true}}$ as the response variable and $U^{\text{approx}}$ as the predictor variable.
Pearson's correlation is used as $U^{\text{true}}_{f,i,t}$ and $U^{\text{approx}}_{f,i,t}$ should be linearly related. Additionally, Pearson's correlation detects skewness and outliers in distribution. As such, 
Pearson's correlation coefficient is a statistical measure of the strength of a linear relationship between paired data \cite{benesty2009pearson}.
For feeder $f$, the Pearson's coefficient is calculated as
\begin{equation}
    \rho_f = \frac{
    \sum_i \sum_t ( U^{\text{true}}_{f,i,t}  - \bar{U}^{\text{true}}_{f,t}) ( U^{\text{approx}}_{f,i,t}  - \bar{U}^{\text{approx}}_{f,t})
    }{\Big(\sum_i \sum_t ( U^{\text{true}}_{f,i,t}  - \bar{U}^{\text{true}}_{f,t})^2 \sum_i \sum_t  ( U^{\text{approx}}_{f,i,t}  - \bar{U}^{\text{approx}}_{f,t})^2\Big)^{1/2}},
    \label{eq:pearsonfeeder}
\end{equation}
where $\bar{U}^{\text{true}}_{f,t}$ and $\bar{U}^{\text{approx}}_{f,t}$ denotes the mean value of true and approximate VUF for feeder $f$ and time instance $t$ respectively.

The Pearson coefficient for all feeders denoted as $\rho$ is the mean value of $\rho_f$ for all feeders, 
\begin{equation}
    \rho = \frac{1}{F}\sum_{f=1}^F \rho_f.
    \label{eq:meanPearson}
\end{equation}
The coefficient of determination suggests $\rho^2$ of the variability of ${U}^{\text{true}}$ is explained by ${U}^{\text{approx}}$. It is given as
\begin{equation}
    \rho^2 = \frac{1}{F} \sum_{f=1}^F \rho_f^2.
\end{equation}
Next, we present three numerical case studies where a comparison between the true and approximate VUF definitions is made. The approximate VUF calculation has the potential to be applied in real-life LV DN, as most DN measurement infrastructure devices are magnitude-based. However, it is crucial to quantify the limitation of the approximate VUF definition. 

\section{Numerical case-studies}
\label{sec:section3}
The DN considered is the set of feeders described in \cite{koirala2020non}.
The reduced $3\times 3$ primitive impedance matrix is used to include the effect of an isolated neutral conductor by the reduction proposed in~\cite{koirala2019impedance} to represent a 4-wire European DN by its 3-wire equivalent.
Actual smart meter measurements of hourly energy consumption for one day was considered in the analysis.
Using network information and load profiles, three-phase multi-period power flows for one day are performed using PowerModelsDistribution.jl in Julia / JuMP
\cite{FOBES2020106664}.

Next, we detail three case studies to provide insights into the comparison of unbalance metrics.

\subsection{Case study 1: Temporal \& locational VUF comparison}
\begin{figure}[!htbp]
	\center
	\includegraphics[width=5.2in]{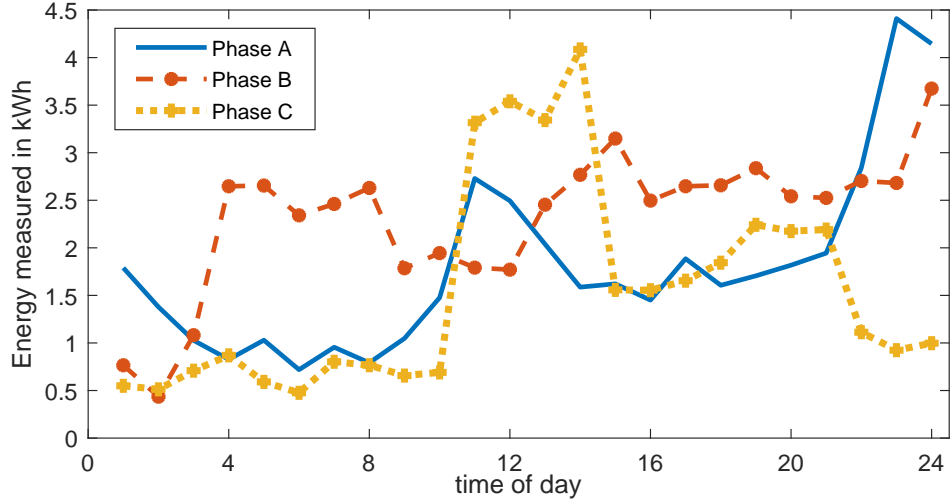}
	\caption{\small{Per-phase energy measured in feeder head. }}
	\label{fig:power}
\end{figure}

\begin{figure}[!htbp]
	\center
	\includegraphics[width=4.5in]{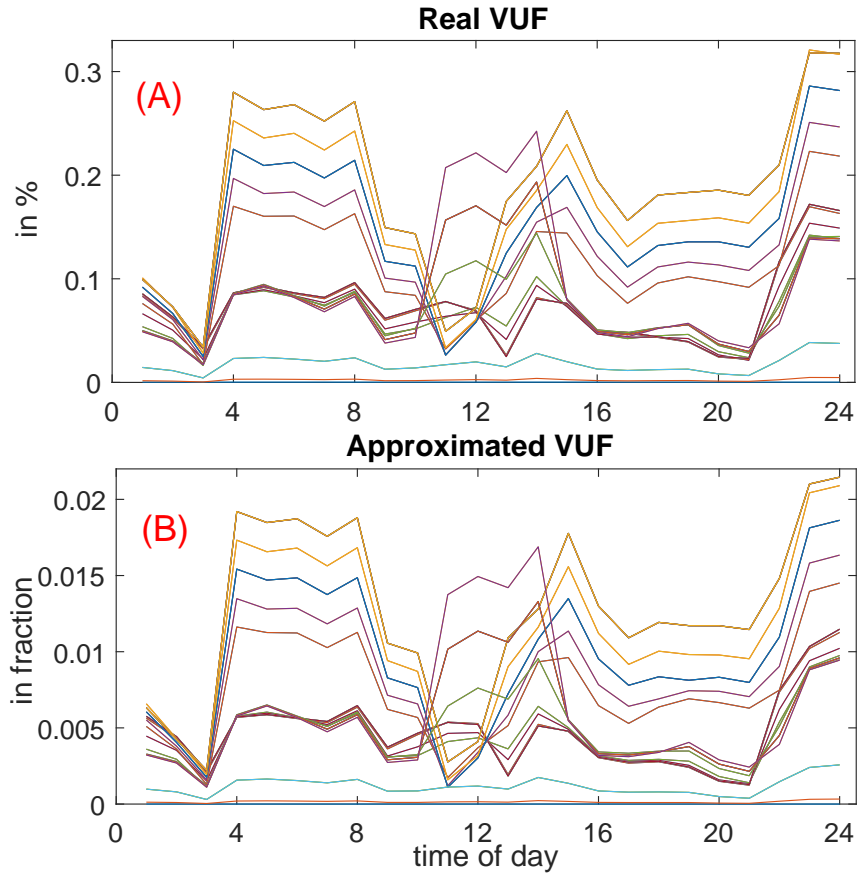}
	\caption{\small{VUF temporal variation over a day; each line denotes a node of 41 bus DN: (a) true VUF in \%, (b) approximate VUF in fraction. }}
	\label{fig:compare}
\end{figure}

\begin{figure}[!htbp]
	\center
	\includegraphics[width=4.5in]{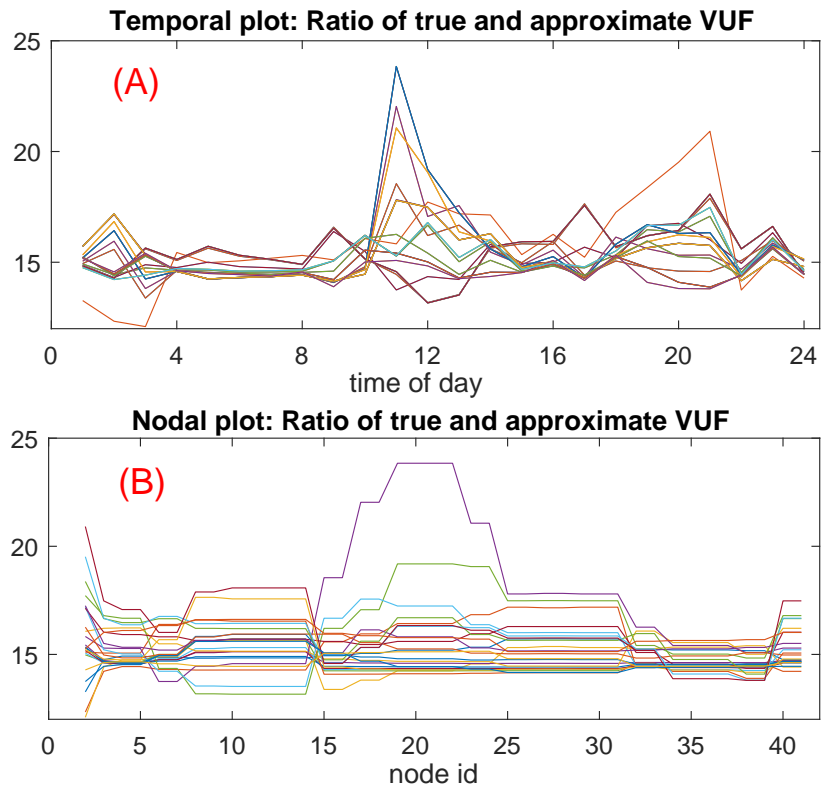}
	\caption{\small{{The ratio of true and approximate VUF: (a) temporal, (b) nodal.}}}
	\label{fig:ratio}
\end{figure}

A typical LV feeder is analysed for assessing the VUF metrics based on the location of a node in a network and time of day.
The LV feeder selected for temporal and locational VUF metric comparison has an R/X ratio of $\approx 6.8$\footnote{Cumulative R/X ratio of DN is calculated by taking the ratio of the sum of all line resistance over the sum of all line reactance. DN is composed of different line parameters with varying R/X ratio. In order to have the analysis tractable, the cumulative R/X of a DN is taken.}. This value corresponds to a typical LV feeder.
This feeder consists of 41 nodes and 18 loads are connected. 16 of these loads are single-phase, and 2 three-phase loads are connected.
The aggregate phase load measured at the feeder head is shown in Fig.~\ref{fig:power}.
Observe that the loading of the three phases is not balanced, resulting in voltage unbalance.
The VUF calculated using \eqref{eq:truedef} and \eqref{eq:approxdef} are shown in Fig.~\ref{fig:compare}. 
{
Observe that in Fig.~\ref{fig:compare}, the approximate definition accurately captures the variation of the true VUF, however, the magnitude of the VUFs are not comparable. The ratio of the true VUF and the approximate VUF remains stable at around 15 for this DN.}
The ratio between the true and approximate VUF is largest when the DN voltages are reasonably balanced, leading to a numerical singularity, see Fig. \ref{fig:ratio}. It can be said that under a fairly balanced DN, the approximate VUF definition will provide an under-estimate of the actual DN unbalance.

\begin{figure}[!htbp]
	\center
	\includegraphics[width=4.8in]{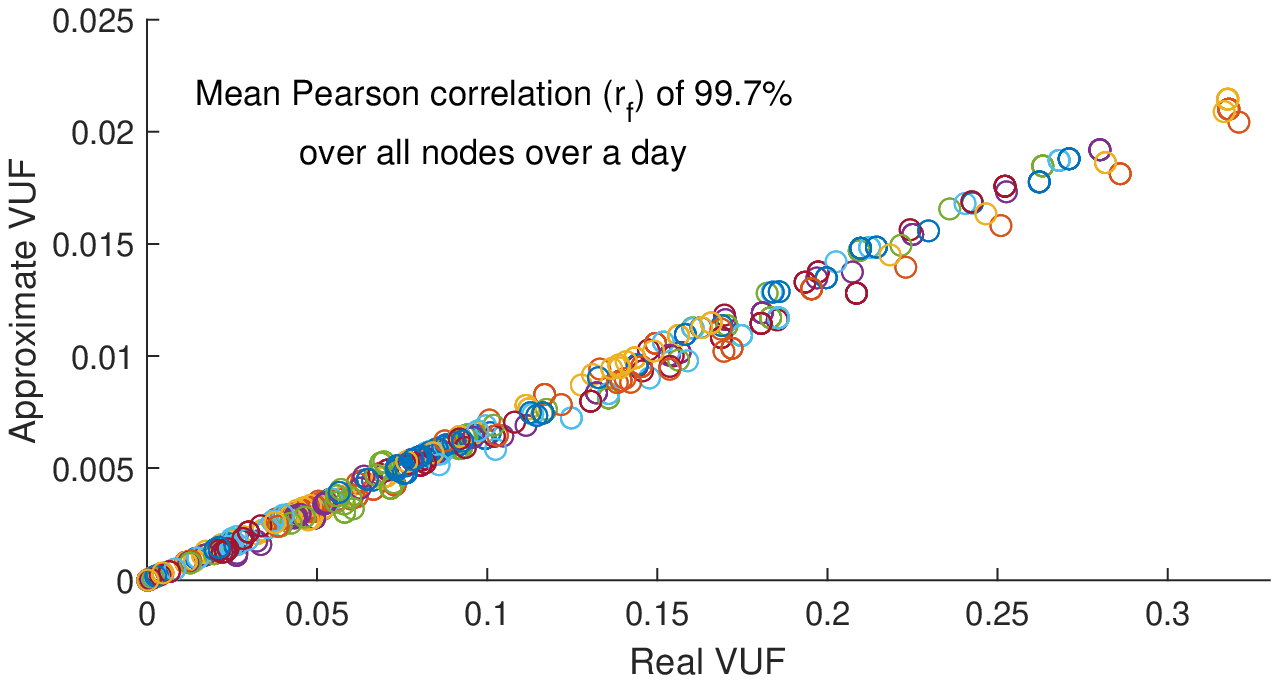}
	\vspace{-8pt}
	\caption{\small{Scatter plot of real and approximate VUF}}
	\label{fig:compare3}
\end{figure}
{
Figure \ref{fig:ratio}(A) and (B) shows the ratio of true and approximate VUF plotter over time and over nodal id respectively.
It can be observed that the ratio of true and approximate VUF remains fairly constant except for instances where load changes happened and/or DN unbalance is very small, causing numerical aberrations.}
Fig.~\ref{fig:compare3} shows the scatter plot of true and approximate VUF calculations. Observe the linear relationship between these two definitions. Using
\eqref{eq:pearsonfeeder} we can calculate the Pearson correlation of the true and approximate VUF of DN over all time instances and all DN nodes for the feeder. 
We observe a very high Pearson correlation of $>99.7\%$. This result shows that for this DN, due to the strong linear correlationship, the approximate VUF can be used for indicating DN unbalance.
Next, we perform a similar analysis for 161 DN feeders with varying R/X values.

\subsection{Case study 2: Comparing metrics for unity power factor}
In order to generalize the observations made in case study 1, {a similar analysis for a larger number of DN feeders is performed. The 161 actual DN feeders from a Spanish city~\cite{koirala2020non} is used for comparing true and approximate VUF definitions.}

\begin{figure}[!htbp]
	\center
	\includegraphics[width=4.9in]{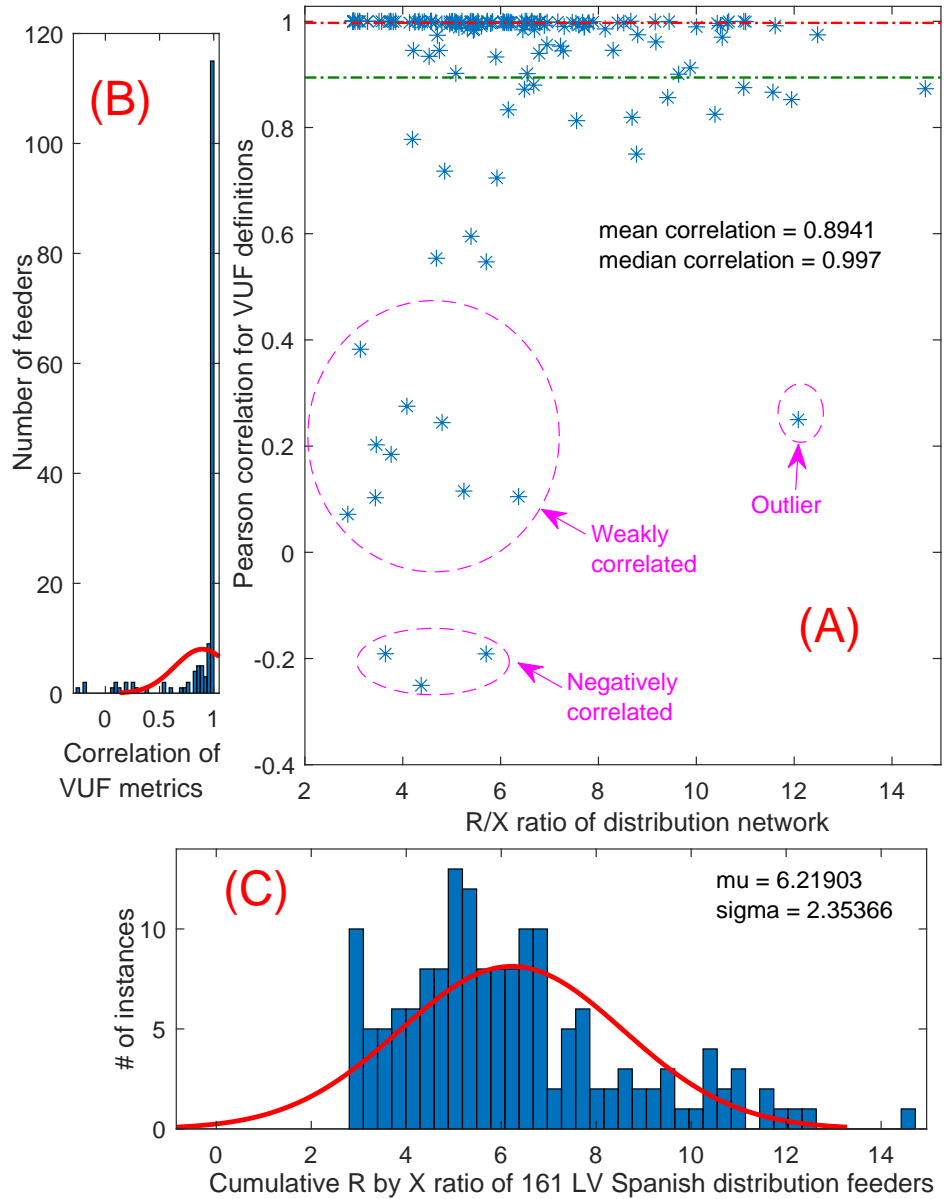}
	\caption{\small{Scatter plot of real and approximate VUF for unity power factor of DN. (a) shows the Pearson correlation of two VUF definitions calculated using \eqref{eq:meanPearson}; each point denotes 1 DN feeder, (b) the distribution of Pearson correlation of DN feeders is shown, (c) the distribution of R/X of 161 DN feeders is shown}}
	\label{fig:unity}
\end{figure}

Fig.~\ref{fig:unity} summarizes the results for the 161 Spanish feeders. Fig.~\ref{fig:unity}(A) shows the distribution of Pearson correlation on the y-axis, $\rho_f \forall f\in\{1,..,161\}$. On the x-axis, the R/X ratio of the DN is indicated. The mean correlation, $\rho$, for all feeders is 89.41\%. The median, however, is significantly higher and exceeds 99.7\%. The distribution of correlation is shown in Fig.~\ref{fig:unity}(B). Note that 117 of the DN feeders have a correlation close to 1. All feeders with low Pearson correlation, except the point indicated as an outlier in Fig.~\ref{fig:unity}(A), have a low R/X value of the feeder. This indicates that for feeders with a high R/X value, the approximate VUF definition emulates the true VUF definition.
Fig.~\ref{fig:unity}(C) shows the distribution of R/X ratio of the DN. The mean value of this distribution is 6.2 with a variance of 2.35.
In Fig.~\ref{fig:unity}(A), the low correlated DN feeders can be clustered into three 
categories: (i) outlier, (ii) negatively correlated consisting of 3 feeders and (iii) weakly correlated consisting of 9 feeders.

\begin{table}
\small
\centering
\caption{\small{VUF metric comparison for DN feeders}}
\begin{tabular}{c|c|c|c|c|c|c} 
\begin{tabular}[c]{@{}l@{}}Feeder \\category\end{tabular}                        & R       & X      & R/X   & $\rho$ & \begin{tabular}[c]{@{}l@{}}Mean \\$U^{\text{true}}_f$\end{tabular} & \begin{tabular}[c]{@{}l@{}}Mean \\$U^{\text{approx}}_f$\end{tabular}  \\ 
\hline
\hline
Outlier                                                                          & 0.074   & 0.006  & 12.08 & 0.25        & \multirow{13}{*}{\begin{tabular}[c]{@{}l@{}}$\approx$\\0.00028 \\ \% \end{tabular}}  & 1e-12                                                      \\ 
\cline{1-5}\cline{7-7}
\multirow{3}{*}{\begin{tabular}[c]{@{}l@{}}Negatively \\correlated\end{tabular}} & 0.222   & 0.061  & 3.64  & -0.19       &                                                                                                                & 8.1e-10                                                    \\ 
\cline{2-5}\cline{7-7}
                                                                                 & 0.353   & 0.619  & 5.71  & -0.19       &                                                                                                                & 1.7e-11                                                    \\ 
\cline{2-5}\cline{7-7}
                                                                                 & 1.022   & 0.234  & 4.38  & -0.25       &                                                                                                                & 6.9e-11                                                    \\ 
\cline{1-5}\cline{7-7}
\multirow{9}{*}{\begin{tabular}[c]{@{}l@{}}Weakly \\correlated\end{tabular}}     & 0.171   & 0.059  & 2.88  & 0.07        &                                                                                                                & 5.1e-11             \\ 
\cline{2-5}\cline{7-7}
                                                                                 & 0.086   & 0.027~ & 3.13  & 0.38        &                                                                                                                & 9.9e-12 \\ 
\cline{2-5}\cline{7-7}
                                                                                 & 0.293   & 0.085  & 3.44  & 0.10        &                                                                                                                & 2.2e-11                                                        \\ 
\cline{2-5}\cline{7-7}
                                                                                 & 0.265   & 0.077  & 3.45  & 0.20        &                                                                                                                & 2.8e-11                                                  \\ 
\cline{2-5}\cline{7-7}
                                                                                 & 0.308 ~ & 0.082  & 3.76  & 0.18        &                                                                                                                & 2.2e-9                                                     \\ 
\cline{2-5}\cline{7-7}
                                                                                 & 0.224   & 0.055  & 4.08  & 0.28        &                                                                                                                & 1.3e-11                                                    \\ 
\cline{2-5}\cline{7-7}
                                                                                 & 0.138   & 0.029  & 4.80  & 0.24        &                                                                                                                & 3.9e-10                                                    \\ 
\cline{2-5}\cline{7-7}
                                                                                 & 0.543   & 0.104  & 5.25  & 0.11        &                                                                                                                & 5.7e-10                                                    \\ 
\cline{2-5}\cline{7-7}
                                                                                 & 2.388   & 0.375  & 6.36  & 0.10        &                                                                                                                & 9.3e-11                                                    \\
\hline
\end{tabular}
\label{tab:comp}
\end{table}
The outlier feeder has a single three-phase consumer and low resistance and high R/X ratio, refer to Table \ref{tab:comp}. The SM measures the cumulative energy in three phases. It is assumed that the load is split equally among the phases. In such a scenario, the feeder is perfectly balanced and the VUF indices do not work as expected {due to numerical singularity}. 
Table \ref{tab:comp} also shows feeder details for weakly and negatively correlated feeders. Observe that all these feeders have very small amounts of load unbalance. The absence of unbalance is either due to the connection of solely three-phase consumers or customers being distributed over the three phases, achieving an almost balanced feeder loading. 
We can observe that approximate phase magnitude-based VUF is an under-estimate of true VUF in such balanced conditions. 
\begin{table}[!htbp]
\small
\caption{\small{VUF with varying power factor}}
\centering
\begin{tabular}{c|c|c|c|c}
Power factor   & 1     & 0.9   & 0.8   & 0.7    \\ \hline \hline
Mean VUF \%    & 0.038 & 0.043 & 0.048 & 0.556  \\
Max VUF \%     & 0.929 & 1.075 & 1.252 & 1.485  \\
Correlation $\rho$ \% & 89.4  & 90.6  & 90.7  & 91.1  \\ 
$\rho^2$ \% & 79.92  & 82.1  & 82.3  & 83.0  \\\hline
\end{tabular}
    \label{tab:vuf}
\end{table}
\subsection{Case study 3: Comparing metrics for varying power factor}
{This case study assesses the impact of the power factor on variation of VUF metrics. }
Observe from Table \ref{tab:vuf} that deterioration of DN power factor does not impact the Pearson correlation coefficient $\rho$, and coefficient of determination $\rho^2$.
For this numerical case study, the power factor of all loads is assumed to be the same.
Due to the decrease in the DN power factor, the unbalance increases slightly. The increase in VUF makes the approximate VUF definition {somewhat} more correlated to the true VUF. Thus, we can conclude that the deteriorating power factor does not impact the accuracy of the approximate VUF. Further, with increasing DN unbalance, the approximate VUF is better correlated with true VUF.

\section{Conclusions and future work}
\label{sec:section4}

{Distribution networks (DN) have inevitable voltage imbalance due to the nature of the connected generation and load. With the ongoing energy transition, we see an increase in large and variable single phase loads (such as EV) and generation (DG). As most current digital meters only provide voltage magnitude measurements, this paper evaluates whether using the voltage magnitude to calculate the approximate voltage unbalance factor (VUF) can be an acceptable proxy for the phasor based true VUF. The evaluation is done on a large number of real-world feeders.}
{We observe that the Pearson correlation for approximate and true VUF metrics have a median exceeding 99.7\% and a mean of 89.4\% for the analysed feeders. We further observe that
the approximate VUF provides a slight underestimation of true VUF in highly balanced DN feeders.}
{The key observation of this work is that the approximate VUF can reliably be applied on DN with a high R/X value. Although these two metrics are not the same in magnitude and an appropriate scaling will be needed for a fair comparison.}
{However, further evaluation is needed to quantify the impact of the R/X ratio on the varying correlation of true and approximate VUF definitions
and to better understand the impact of load profiles on DN unbalance and the projection of these parameter variations on unbalance metrics.}

\section*{Acknowledgement}
This work is supported by the H2020 EUniversal project, grant agreement ID: 864334 (\url{https://euniversal.eu/}).

\bibliographystyle{IEEEtran}
\bibliography{reference}

\end{document}